\documentclass[twocolumn]{aastex6}
\let\pwifjournal=\iffalse

\usepackage[T1]{fontenc}
\pwifjournal\else
  \usepackage{microtype}
\fi

\pwifjournal\else
  \makeatletter
  \renewcommand\plotone[1]{%
    \centering \leavevmode \setlength{\plot@width}{0.99\linewidth}
    \includegraphics[width={\eps@scaling\plot@width}]{#1}%
  }%
  \makeatother
\fi

\pwifjournal\else
  \usepackage{etoolbox}
  \makeatletter
  \patchcmd{\NAT@citex}
    {\@citea\NAT@hyper@{%
       \NAT@nmfmt{\NAT@nm}%
       \hyper@natlinkbreak{\NAT@aysep\NAT@spacechar}{\@citeb\@extra@b@citeb}%
       \NAT@date}}
    {\@citea\NAT@nmfmt{\NAT@nm}%
     \NAT@aysep\NAT@spacechar\NAT@hyper@{\NAT@date}}{}{}
  \patchcmd{\NAT@citex}
    {\@citea\NAT@hyper@{%
       \NAT@nmfmt{\NAT@nm}%
       \hyper@natlinkbreak{\NAT@spacechar\NAT@@open\if*#1*\else#1\NAT@spacechar\fi}%
         {\@citeb\@extra@b@citeb}%
       \NAT@date}}
    {\@citea\NAT@nmfmt{\NAT@nm}%
     \NAT@spacechar\NAT@@open\if*#1*\else#1\NAT@spacechar\fi\NAT@hyper@{\NAT@date}}
    {}{}
  \makeatother
\fi

\usepackage{natbib}
\makeatletter
\newcommand\pkgw@simpfx{http://simbad.u-strasbg.fr/simbad/sim-id?Ident=}
\newcommand\MakeObj[4][\@empty]{
  \pwifjournal%
    \expandafter\newcommand\csname pkgwobj@c@#2\endcsname[1]{\protect\object[#4]{##1}}%
  \else%
    \expandafter\newcommand\csname pkgwobj@c@#2\endcsname[1]{\href{\pkgw@simpfx #3}{##1}}%
  \fi%
  \expandafter\newcommand\csname pkgwobj@f#2\endcsname{#4}%
  \ifx\@empty#1%
    \expandafter\newcommand\csname pkgwobj@s#2\endcsname{#4}%
  \else%
    \expandafter\newcommand\csname pkgwobj@s#2\endcsname{#1}%
  \fi}%
\newcommand{\obj}[1]{%
  \expandafter\ifx\csname pkgwobj@c@#1\endcsname\relax%
    \textbf{[unknown object!]}%
  \else%
    \csname pkgwobj@c@#1\endcsname{\csname pkgwobj@s#1\endcsname}%
  \fi}
\newcommand{\objf}[1]{%
  \expandafter\ifx\csname pkgwobj@c@#1\endcsname\relax%
    \textbf{[unknown object!]}%
  \else%
    \csname pkgwobj@c@#1\endcsname{\csname pkgwobj@f#1\endcsname}%
  \fi}
\newcommand\OverrideObjectURL[2]{
  \expandafter\renewcommand\csname pkgwobj@c@#1\endcsname[1]{\href{#2}{##1}}}%
\makeatother

\MakeObj[WISE~0716$-$19]{0716-19}{temp}{WISE~J071634.59$-$190039.2}
\OverrideObjectURL{0716-19}{http://vizier.u-strasbg.fr/viz-bin/VizieR-5?-ref=VIZ56cf52489d33%
  &-out.add=.&-source=II/328/allwise&AllWISE===J071634.59-190039.2}

\newcommand\apx{\ensuremath{\sim}}
\newcommand\citeeg[1]{\citep[\emph{e.g.},][]{#1}}
\renewcommand\deg{\ensuremath{^\circ}}
\newcommand\dm[1]{\ensuremath{{\rm DM}_{\rm #1}}}
\newcommand\dmu{cm$^{-3}$~pc}
\newcommand\dt{\ensuremath{{\Delta t}}}

\newcommand\ha{\ensuremath{{\text{H}\alpha}}}

\newcommand\msun{\ensuremath{\text{M}_\odot}}
\newcommand\speclum{erg~s$^{-1}$~Hz$^{-1}$}
\newcommand\thefrb{FRB~150418}

\begin{document}

\title{No precise localization for \thefrb: claimed radio transient is AGN
  variability}
\author{
  P.~K.~G. Williams,
  E. Berger
}
\affil{Harvard-Smithsonian Center for Astrophysics, 60 Garden Street,
  Cambridge, MA 02138, USA}
\email{pwilliams@cfa.harvard.edu}

\slugcomment{Resubmitted to ApJL}
\shorttitle{No precise localization for \thefrb}
\shortauthors{Williams \& Berger}

\begin{abstract}
  Keane et al. have recently claimed to have obtained the first precise
  localization for a Fast Radio Burst (FRB) thanks to the identification of a
  contemporaneous fading slow (\apx week-timescale) radio transient. They use
  this localization to pinpoint the FRB to a galaxy at $z \approx 0.49$ that
  exhibits no discernible star formation activity. We argue that the transient
  is not genuine and that the host candidate, WISE~J071634.59$-$190039.2, is
  instead a radio variable: the available data did not exclude this
  possibility; a random radio variable consistent with the observations is not
  unlikely to have a redshift compatible with the FRB dispersion measure; and
  the proposed transient light curve is better explained as a scintillating
  steady source, perhaps also showing an active galactic nucleus (AGN) flare,
  than a synchrotron-emitting blastwave. The radio luminosity of the host
  candidate implies that it is an AGN and we present new late-time Very Large
  Array observations showing that the galaxy is indeed variable at a level
  consistent with the claimed transient. Therefore the claimed precise
  localization and redshift determination for \thefrb\ cannot be justified.
\end{abstract}


\section{Introduction}

The origin of Fast Radio Bursts \citep[FRBs;][]{lbm+07} remains unknown, with
both Galactic and extragalactic scenarios proposed \citeeg{fr14, lsm14, z14}.
The SUrvey for Pulsars and Extragalactic Radio Bursts (SUPERB) project has
recently claimed to have obtained the first precise localization for an FRB by
identification of an associated radio transient that faded over the course of
six days \citep{kjb+16}. This transient was located in a seemingly passive
elliptical galaxy at $z = 0.492 \pm 0.008$, a phenomenology which they argued
to be consistent with the possible origin of (at least some) FRBs in compact
object mergers \citeeg{z14}. This would be a truly exciting discovery,
confirming the cosmological origin of (at least some) FRBs and hence also
their extreme physics, their utility as a probe of the intergalactic medium
\citeeg{m14}, and the possibility that FRBs may be prompt, localizable
electromagnetic tracers of gravitational-wave events \citep{aaa+16}. The
claimed localization of \thefrb\ has already been used to investigate the
properties of its progenitor system \citep{x.wyw+16, x.z16} and place limits
on the equivalence principle \citep{x.tk16} and the mass of the photon
\citep{x.bem+16}.

Here we argue that the properties of the long-term radio emission from the
proposed host point to a different and more mundane interpretation: that the
observed variable radio emission is instead due to AGN activity, and that the
variable emission and galaxy are not necessarily related to \thefrb. Reasons
to doubt the association (\autoref{s.doubts}) include failure to exclude
variable radio emission as a potential origin of the signal and the
disagreement between the proposed transient light curve and synchrotron
blastwave models, which are used to describe all confirmed classes of
extragalatic radio transients. We also show that the agreement between the
host candidate redshift and the dispersion measure (DM) of \thefrb\ is not
unlikely, if the host candidate was selected based on short-timescale radio
variability. We argue that the host candidate's quiescent radio luminosity
implies that it hosts an AGN (\autoref{s.agn}) and present new data that we
obtained with the Karl G. Jansky Very Large Array (VLA) demonstrating that it
is indeed a variable radio source, attaining flux densities comparable to
those attributed to the proposed radio transient (\autoref{s.vla}). In
\autoref{s.conc} we conclude that while other lines of evidence suggest an
extragalactic origin for at least some FRBs \citep{mls+15}, that currently
available for \thefrb\ is unpersuasive.

The \thefrb\ host galaxy candidate is robustly detected in AllWISE imagery and
is cataloged in the AllWISE Data Release as \objf{0716-19}. Hereafter we refer
to it as \obj{0716-19}.

\section{Reasons to doubt association of \thefrb\ and \obj{0716-19}}
\label{s.doubts}

\begin{figure}
  \centering
  \includegraphics[width=\linewidth]{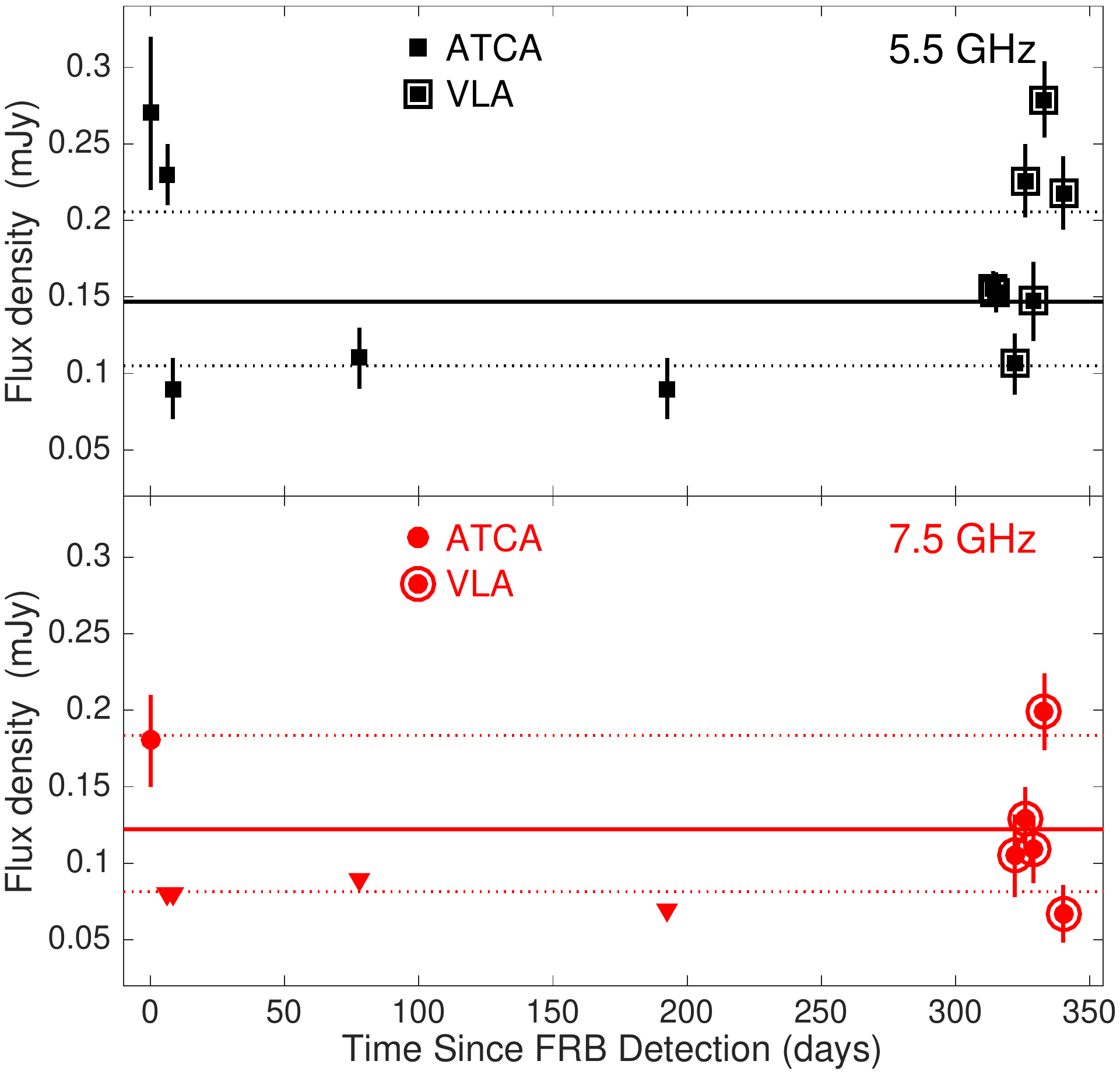}
  \caption{Radio light curve of \obj{0716-19} at 5.5 and 7.5~GHz (black and
    red, respectively) from VLA and ATCA (points with and without outlines,
    respectively). The ATCA data are from \citet{kjb+16}. Each panel shows the
    best-fit model of a steady source affected by scintillation
    (\autoref{s.models}), with the dotted lines showing the range of flux
    variation expected from refractive scintillation. The first two VLA epochs
    did not obtain data at 7.5~GHz and have been averaged together for
    clarity.}
  \label{f.alldata}
\end{figure}

\citet{kjb+16} followed up the detection of \thefrb\ with radio observations
at several frequencies using several different telescopes. They achieved five
detections of \obj{0716-19} with the Australia Telescope Compact Array (ATCA)
at 5.5~GHz and one at 7.5~GHz; observations at other frequencies resulted in
nondetections. We reproduce the ATCA data in \autoref{f.alldata} using the
measurements provided in Extended~Data~Table~1 of \citet{kjb+16}, combining
them with our new observations (\autoref{s.vla}). Here and below, we take the
time of each observation to be its midpoint as computed by offsetting its
tabulated start time by half of its duration. We compute $\dt = \text{MJD} -
57130.19$ to express the approximate time after \thefrb\ in days. We do not
apply barycentric or timescale corrections, which are not relevant to our
analysis. There is a discrepancy between the \citet{kjb+16} table and their
Figure~2: the table incorrectly lists the fourth ATCA epoch as occurring on
2015~June~4 when it should be 2015~July~4 (S.~Johnston, 2016, priv. comm.).

In this section we provide several \textit{a priori} reasons to doubt the
association between \thefrb\ and \obj{0716-19}.

\bigskip\bigskip 

\subsection{Failure to exclude coincident variable source}
\label{s.stats}

The analysis of \citet{kjb+16} examines the probability of the chance
discovery of an unassociated radio transient in their search field, but not
the probability of the chance discovery of a variable radio source. The odds
of the latter are non-negligible, as implied by the presence of a second
compact variable radio source within the Parkes beam \citep{kjb+16}. The five
detections of the ATCA light curve of \obj{0716-19} are insufficient to reject
the possibility that it is a variable radio source, as demonstrated
empirically by our new data showing that it in fact is one (\autoref{s.vla}).

Precise statements regarding the probability of chance detection of a
candidate matching the characteristics reported by \citet{kjb+16} cannot be
made without information regarding the total number of FRB localization
regions searched by the SUPERB project and the process by which candidate
transients were filtered, which is not currently available. However, in a
catalog of 3652 compact sources brighter than \apx0.1~mJy at 3~GHz produced
for the Caltech-NRAO Stripe~82 Survey pilot (CNSSp), \citet{mhb+16} find that
$3.9^{+0.5}_{-0.9}$\% of them are variable at the $>$30\% level. They only
classified two sources as transients, implying that variables outnumber
transients by a factor of $\approx$70 and that the ``headline'' chance
coincidence probability of $<$0.1\% reported by \citet{kjb+16} may be
underestimated by a comparable amount. More generally, studies in which the
analysis performed depends on the data taken will inevitably yield
overconfident significance metrics due to the ``garden of forking paths''
effect \citep{gl14}.

Furthermore, the probability of a radio variable masquerading as a radio
transient in the particular data set reported by \citet{kjb+16} may be even
higher. \citet{ofb+11} used the VLA to search a total area of 2.66~deg$^2$ for
radio transients and variables. They find that 30\% (30 out of 98) of sources
brighter than 1.5~mJy at 5~GHz are variable at the 4$\sigma$ level. Given the
three radio sources with $S_\nu \gtrsim 0.1$~mJy detected in our VLA imaging
(\autoref{s.vla}), which has a position and total area close to that of the
Parkes search area, the expected number of radio variables in the field is
therefore of order unity. Using the deep 5~GHz source counts of
\citet{fwkk91}, \apx16 sources brighter than 0.1~mJy are expected to be found
in each Parkes beam. The typical Parkes search area may therefore host
multiple variable radio sources.

\citet{ofb+11} note that the rate of variables found in their survey is higher
than comparable surveys and attribute this to their choice of observing
frequency, the short averaging times of their observations, and the low
Galactic latitude ($b \apx 6$--8\deg) of their survey. All of these factors
apply to the observations of \citet{kjb+16}, with \obj{0716-19} being found at
$b \apx -3.2$\deg. The correlation between low Galactic latitude and increased
incidence of variability is well established and is due at least in part to
higher levels of refractive scintillation through the denser ISM
\citep[\autoref{s.models};][]{sfgp89, r90}, implying that the increase in the
number of variable sources is not only due to foreground objects.

\subsection{Significance of host galaxy redshift}
\label{s.mcmc}

It may be argued that the agreement between the redshift of \obj{0716-19} and
the DM of \thefrb, given standard cosmological assumptions, supports the
conclusion that the two are associated. Here we demonstrate that consistency
between these is not unlikely even if the FRB and galaxy are unrelated.

We performed a Markov Chain Monte Carlo (MCMC) simulation to characterize the
host galaxy redshifts that would have been found to be consistent with the DM
of \thefrb, given the assumptions made by \citet{kjb+16}\footnote{We take this
  approach, rather than considering the likelihood that the the FRB DM would
  be found to be consistent with the host galaxy redshift, because the latter
  approach requires assumptions about the underlying distribution of FRB DMs,
  which is not well-constrained, as well as speculation as to what DM model
  would have been adopted by \citet{kjb+16} had a different DM been
  measured.}. The parameters are summarized in \autoref{t.mcmc}; the model is
defined by Equation~1 and the surrounding discussion in \citet{kjb+16}. We add
a small (1\%) uncertainty on the fraction of baryons contained in the
intergalactic medium (IGM). Using a likelihood defined by the measured FRB DM
and the priors listed in \autoref{t.mcmc}, we sampled from the posterior using
the \textsf{emcee} package \citep{the.emcee}, which implements the
\citet{gw10} affine-invariant sampling algorithm. We used 256 walkers divided
into 8 independent groups each taking 8192 steps, thinning by a factor of 16
and discarding the first half of the samples from each walker. The mean
proposal acceptance fraction was 41\%, and there were in total \apx8500
independent samples of the redshift $z$, accounting for the estimated chain
autocorrelation length of \apx6 samples after thinning. The $\hat R$
convergence criterion for the redshift parameter reached 1.08, implying good
convergence \citep{the.bda3}.

\begin{figure}
  \centering
  \includegraphics[width=\linewidth]{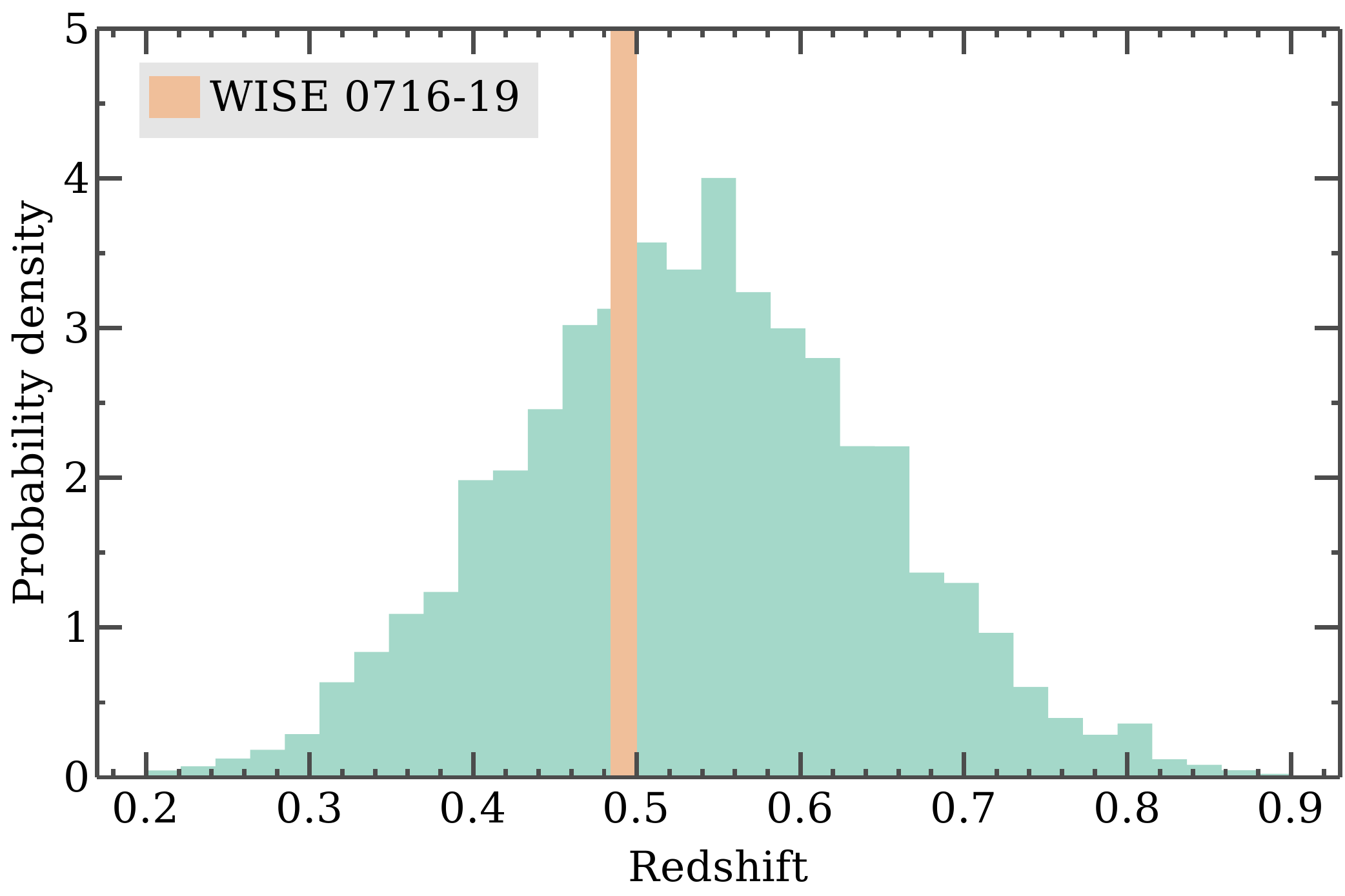}
  \caption{Posterior distribution of redshifts consistent with the observed DM
    of \thefrb, from MCMC analysis using the nearly same model and
    uncertainties as \citet{kjb+16} (\autoref{s.mcmc}). The vertical axis is
    in units of probability per unit redshift so that the area under the curve
    is unity. Host galaxy redshifts in the range 0.42--0.65 are consistent
    with the data at the 1$\sigma$ level, given the FRB DM and adopted model.}
  \label{f.redshifts}
\end{figure}

\autoref{f.redshifts} shows the redshift posterior samples marginalized over
all other parameters. Given the model and data, host galaxy redshifts in the
range 0.42--0.65 can be judged consistent with the measured DM of \thefrb\ at
the 1$\sigma$ level. The true range of host redshifts consistent with the data
is broader than this --- and our analysis is thus conservative --- even if the
assumption of an extragalactic origin is maintained, because other DM models
are valid. For example, if, as we argue, the elliptical galaxy is not
associated with \thefrb\ and the true host is allowed to be a spiral rather
than elliptical galaxy, its DM contribution could be significantly larger than
the value assumed in the present model, broadening the distribution of allowed
redshifts to include lower values.

Radio AGN are generally found at redshifts comparable to those allowed by the
FRB DM measurement \citeeg{ccg+98}. In its unbiased search for radio variables
and transients, the CNSSp discovered 142 such objects in observations at
2--4~GHz. Of the 35 variables with variability timescales less than 1~week,
there are 13 measured redshifts, ranging from 0.15 to 0.84 with a mean of
0.45. We further note that 90\% (32/35) of these variables are classified as
AGN. Of the full sample of CNSSp variables with redshift measurements, 22\%
(15/69) and 41\% (28/69) are within the 1$\sigma$ and 2$\sigma$ limits of the
posterior, respectively. A radio source selected on the basis of its
variability is therefore not unlikely to have a redshift compatible with the
DM of \thefrb.

\begin{deluxetable*}{lccl}
  \tablewidth{0em}
  \tablecaption{Parameters of model used in DM MCMC analysis\label{t.mcmc}}
  \tablehead{
    \colhead{Parameter} &
    \colhead{Symbol} &
    \colhead{Units} &
    \colhead{Prior}
  }
  \startdata
  Host galaxy redshift & $z$ & --- & $U(10^{-4}, 20)$\tablenotemark{a} \\
  Host galaxy DM & \dm{host} & \dmu & $N(37, 37 \times 20\%)$\tablenotemark{b} \\
  Milky way DM & \dm{MW} & \dmu & $N(188.5, 188.5 \times 20\%)$ \\
  Milky way halo DM & \dm{halo} & \dmu & $N(30, 5)$\tablenotemark{c} \\
  Dark energy content of universe & $\Omega_\Lambda$ & --- & $N(0.721, 0.025)$ \\
  Matter content of universe & $\Omega_m$ & --- & $N(0.233, 0.023)$ \\
  Baryonic content of universe & $\Omega_b$ & --- & $N(0.0463, 0.0024)$ \\
  Present-day Hubble parameter & $H_0$ & km s$^{-1}$ Mpc$^{-1}$ & $N(70.0, 2.2)$ \\
  Fraction of baryons in IGM & $f_{\rm IGM}$ & --- & $N(0.90, 0.01)$ \\
  \enddata
  \tablenotetext{a}{Denotes a uniform distribution between the specified
    bounds.}
  \tablenotetext{b}{Denotes a normal distribution with the specified mean
    and standard deviation.}
  \tablenotetext{c}{Uncertainty estimated from Figure~2 of \citet{dgbb15}.}
\end{deluxetable*}

\subsection{Light curve of proposed transient and scintillation}
\label{s.models}

The only confirmed slowly-evolving extragalactic radio transients are
synchrotron-emitting blastwaves, which exhibit a clear relationship between
evolutionary timescale and luminosity \citep{mwb15}. From the observed flux of
$F_\nu(5.5\text{ GHz}) \apx 0.27$~mJy at a mid-point of $\dt = 0.2$, and
assuming expansion at $v \approx c$ we infer a brightness temperature of $T_B
\approx 5\times10^{15}$~K, which clearly requires relativistic expansion, with
an inferred Lorentz factor of $\Gamma \approx 6$ to avoid the inverse Compton
catastrophe limit of $T_B \approx 10^{12}$~K. Thus, if the observed emission
is due to a synchrotron-emitting blastwave, it will obey the basic
relativistic afterglow evolution of GRBs \citep{spn98, sph99, gs02}. The
synchrotron emission model is characterized by three break frequencies ---
self-absorption ($\nu_a$), peak ($\nu_m$) and cooling ($\nu_c$) -- and an
overall flux density normalization ($F_{\nu,m}$). These parameters in turn
determine the physical properties of the blastwave: isotropic kinetic energy
($E_{K,{\rm iso}}$), density ($n$), and fractions of post-shock energy in the
relativistic electrons ($\epsilon_e$) and magnetic fields ($\epsilon_B$). The
power law distribution of the relativistic electrons is further determined by
an index, $p$, such that $N(\gamma) \propto \gamma^{-p}$ at $\gamma \ge
\gamma_m$. This model has been used to study GRB afterglows for the past 15
years.

\begin{figure}
  \centering
  \includegraphics[width=0.95\linewidth]{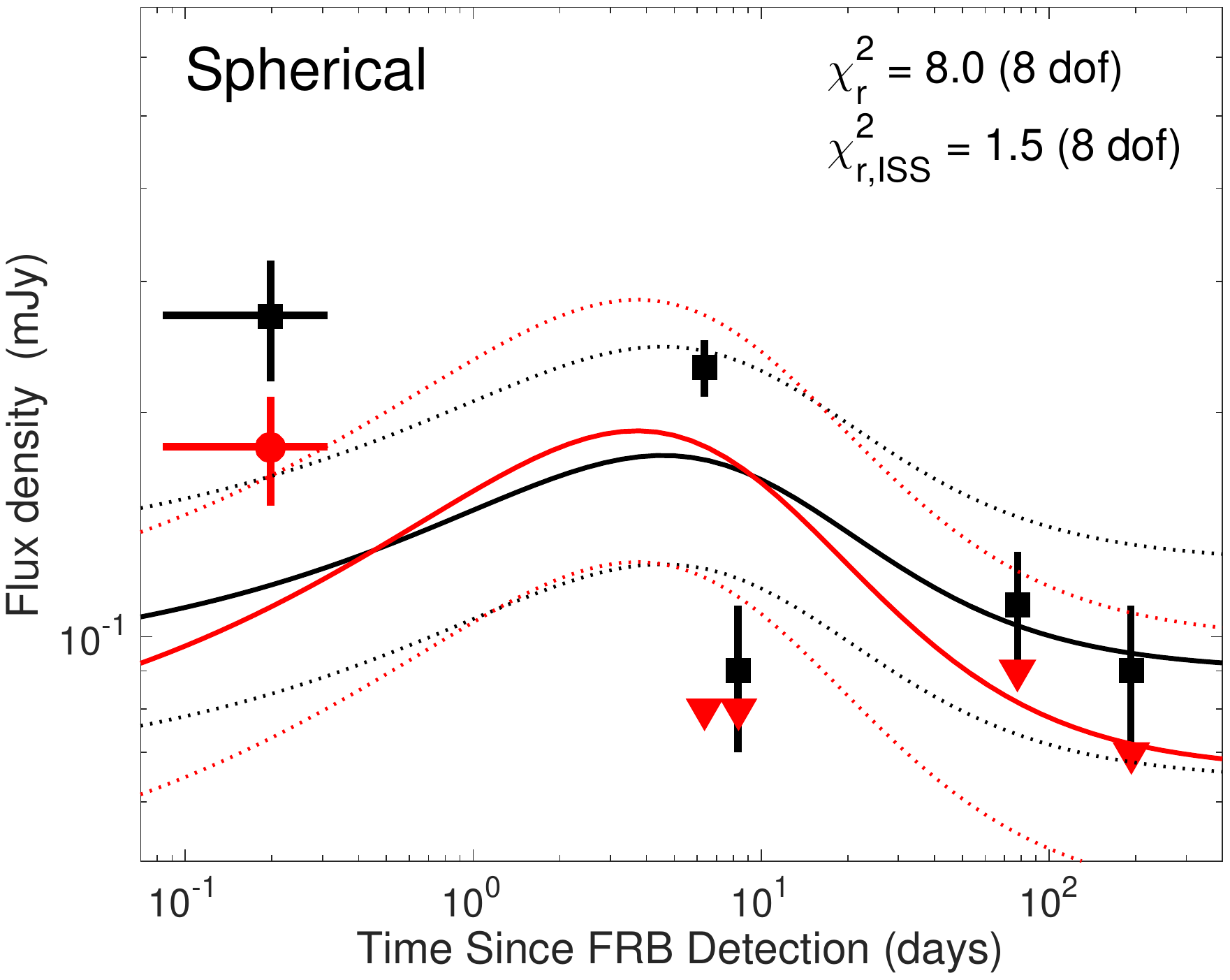}
  \includegraphics[width=0.95\linewidth]{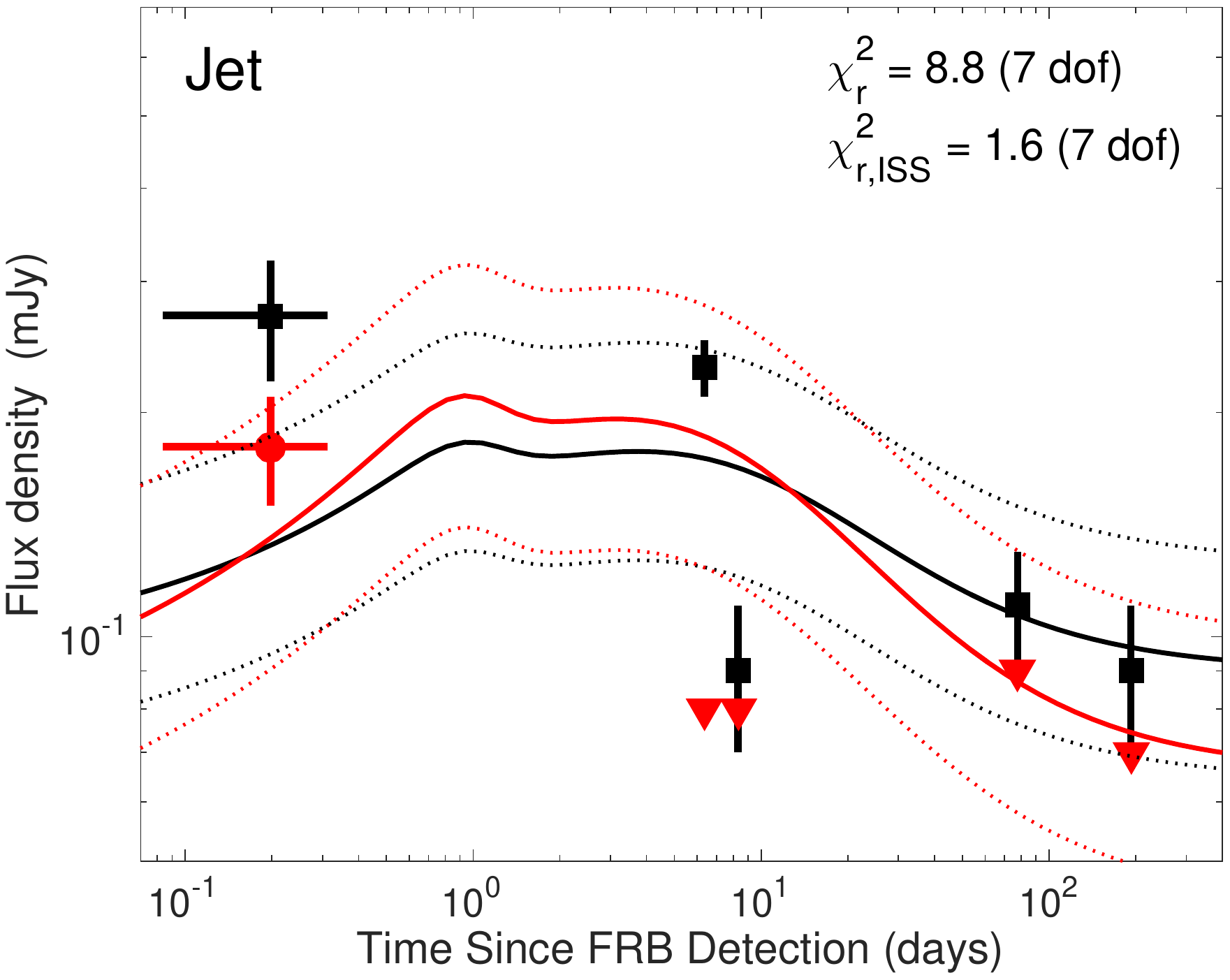}
  \includegraphics[width=0.95\linewidth]{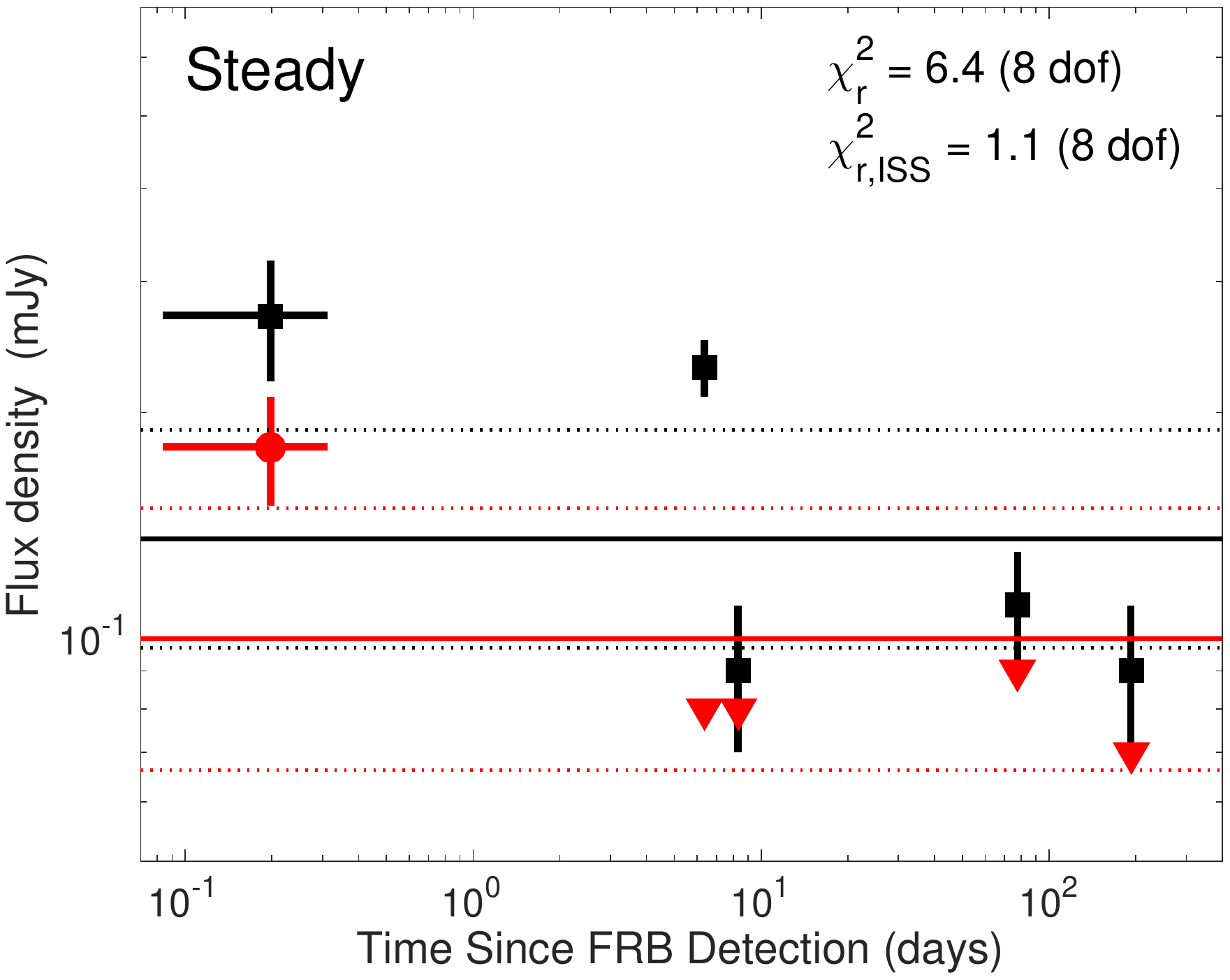}
  \caption{Three theoretical models fit to the ATCA data: synchrotron
    blastwaves with spherical (top) and jetted (middle) geometries, and a
    steady source (bottom). Colors and symbols are as in \autoref{f.alldata},
    with the horizontal error bar on the first point indicating the duration
    of the relevant observation. Regardless of whether RISS is assumed to play
    a role or not, the constant model is more consistent with the data than
    the forward shock afterglow models.}
  \label{f.models}
\end{figure}

Compact radio sources --- including both GRB afterglows and AGN jets --- are
furthermore subject to interstellar scintillation (ISS) by the interstellar
medium of the Milky Way \citep{sfgp89, r90}. For the low Galactic latitude
sight-line to \thefrb\ the scattering measure is large, $\log(\text{SM})
\approx -2.4$ \citep{cl02}, and hence frequencies below $\nu_0 \approx 30$~GHz
are subject to strong scintillation. We do not consider diffractive ISS to be
important because the coherence bandwidth is $\delta\nu / \nu\approx (\nu /
\nu_0)^{17/5} \approx 20$~MHz, much narrower than the GHz bandwidth of the
ATCA and VLA observations. However, strong refractive interstellar
scintillation (RISS) is expected, with a modulation index (rms fractional
variation) of \apx0.4 and 0.5 at 5.5 and 7.5~GHz, respectively. This level of
variability will be present for any source that is compact relative to the
characteristic RISS angular size $\theta_s$. Taking a scattering screen
distance of 1~kpc, we find $\theta_s \apx 50$~$\mu$as at 5.5~GHz \citep{w98},
corresponding to a linear scale of \apx0.2~pc at the redshift of
\obj{0716-19}. In our modeling of the radio emission below we account for the
effect of RISS by adding the expected modulation of the model light curves in
quadrature to the measurements uncertainties.

We model the radio light curve with the afterglow model of \citet{gs02} using
standard parameters for GRB afterglows: $\epsilon_e = 0.1$, $\epsilon_B =
0.01$, and $p = 2.5$. We note that the ATCA data indicate that $\nu_a < 5.5$
GHz, and moreover the radio data do not constrain $\nu_c$, which is typically
located in the optical to X-ray regime. As a result, the model light curves
are degenerate with respect to our choice of $\epsilon_e$ and $\epsilon_B$,
with the inferred values of $E_{K,{\rm iso}}$ and $n$ changing with the choice
of values, but the light curves (and hence the quality of fit) remaining
unchanged. We tested models with values of $\epsilon_e$ and $\epsilon_B$
varying between 0.1 and $10^{-6}$, and find identical $\chi^2_r$ values. We
use models with both a spherical geometry and a jet, leaving the blastwave
kinetic energy and the density as free parameters, as well as the jet opening
angle in the latter model. We also include a constant term with a flux density
of 0.09~mJy at 5.5~GHz and 0.065~mJy at 7.5~GHz to represent the steady
component detected in the ATCA data at $\dt \gtrsim 8$; our 7.5 GHz~steady
component agrees with the ATCA upper limits and assumes a $\nu^{-1}$ spectrum
for this component. The best-fit models are shown in \autoref{f.models}. Both
models provide a poor fit to the data, with $\chi^2_r \approx 8.0$ (spherical;
8 degrees of freedom) and $\approx 8.8$ (jet; 7 degrees of freedom) assuming
no RISS. The inclusion of RISS leads to $\chi^2_r \approx 1.5$ and $\approx
1.6$, respectively. In this latter case, we assume that the steady component
scintillates as well; if it does not (i.e., is not compact), the $\chi^2_r$
values increase since the scintillation-induced uncertainty in the model is
lower.

We next compare these models to a simple steady source which is modulated
purely by RISS. In this case we find a mean flux density of 0.135~mJy at
5.5~GHz and 0.100~mJy at 7.5~GHz (i.e., assuming a $\nu^{-1}$ spectrum). This
simpler model results in $\chi^2_r \approx 6.4$ (8 degrees of freedom) when
ignoring RISS and $\chi^2_r \approx 1.1$ when including RISS.

The synchrotron models are challenged by the data at $\dt < 8$, which show a
rapid evolution in spectral slope that is not expected from a synchrotron
blastwave. More specifically, while the spectral indices of the first and
third epochs of ATCA observations are not atypical, the second epoch implies
an exceptionally steep $\alpha \lesssim -3.4$ between 5.5 and 7.5~GHz. This
variation may be compatible with RISS, which has a correlation bandwidth
$\Delta\nu / \nu \apx 1$. Recent observations suggest that flares in faint AGN
can result in rapid spectral evolution: $\alpha$ evolved from $-1.7$ to $+0.4$
over 15 days in a source (VTC225411$-$010651) found in the CNSSp. We speculate
that this mechanism is at work in this case as well.

Thus, while \citet{kjb+16} (and similarly \citealt{x.z16}) claim that the
post-FRB radio data are consistent with a short GRB afterglow, the data
actually favor other interpretations. The best formal fit to the data is of a
model of a steady source modulated by the inevitable strong refractive
scintillation. The rapid spectral evolution observed at $\dt < 8$ may suggest
the presence of an AGN flare. This spectral evolution is inconsistent with
a synchrotron blastwave, such as a short GRB afterglow.

\section{Alternate interpretation: AGN variability}
\label{s.agn}

In the interpretation of \citet{kjb+16}, the three 5.5~GHz ATCA data points at
$\dt \apx (8, 49, 193)$ are due to quiescent radio emission at a level of
$0.097 \pm 0.012$~mJy, where we have simply taken the weighted mean of the
three measurements. At the redshift of the galaxy this corresponds to a radio
spectral luminosity of \apx$9 \times 10^{29}$~\speclum. Using the standard
relations of \citet{yc02}, the star formation rate (SFR) inferred from the
radio spectral luminosity is \apx$10^2$--$10^3$~\msun~yr$^{-1}$, orders of
magnitude higher than the value of $\leq0.2$~\msun~yr$^{-1}$ that
\citet{kjb+16} infer from \ha\ in the optical spectrum of the galaxy. Thus,
the origin of the quiescent radio emission is not star formation activity.

As argued by \citet{bjfm11} in their investigation of the radio emission from
bright early-type galaxies comparable to \obj{0716-19}, if the galaxy's bright
radio emission is not due to star formation, the alternative source is AGN
activity. This is immediately worrisome because AGN are both intrinsically and
extrinsically variable (\autoref{s.models}) and can thus falsely appear as
transient radio sources. While the spectrum of the host does not show clear
quasar features, spectra of matched SDSS-FIRST sources show that optical
signatures of AGN activity are frequently not visible in spectra of luminous
early-type galaxies with radio emission similar to \obj{0716-19}
\citep{imk+02}. Studies of radio-loud AGN demonstrate that the WISE colors are
consistent with AGN activity \citep{ghj14}.

\section{VLA follow-up observations}
\label{s.vla}

To test the AGN hypothesis, we are obtaining follow-up observations with the
VLA using Director's Discretionary time. Here we present the first results
from our program (number VLA/16A-431).

\begin{deluxetable*}{lllllllll}
  \tablewidth{0em}
  \tabletypesize{\scriptsize}
  \tablecaption{Parameters of VLA observations\label{t.obs}}
  \scriptsize 
  \tablehead{
    \colhead{Parameter} &
    \colhead{Units} &
    \multicolumn{7}{c}{Epoch} \\
    \cline{3-9}
    & &
    \colhead{Feb. 27} &
    \colhead{Feb. 28} &
    \colhead{Mar. 05} &
    \colhead{Mar. 08} &
    \colhead{Mar. 11} &
    \colhead{Mar. 16} &
    \colhead{Mar. 23}
  }
  \startdata
  Observation start time & MJD &
    57445.028 &
    57446.018 &
    57452.015 &
    57456.006 &
    57458.989 &
    57463.972 &
    57470.967 \\
  Observation duration & minutes &
    90 & 90 & 30 & 30 & 30 & 30 & 30 \\[1em]
  5.5 GHz: \\
  ~~~Synthesized beam size & arcsec &
    $8.6 \times 3.3$ &
    $8.9 \times 3.3$ &
    $9.7 \times 3.2$ &
    $9.6 \times 3.2$ &
    $10.4 \times 3.2$ &
    $11.9 \times 3.2$ &
    $9.8 \times 3.6$ \\
  ~~~Calibrator flux density\tablenotemark{a} & mJy &
    $1.197 \pm 0.002$ &
    $1.187 \pm 0.002$ &
    $1.184 \pm 0.004$ &
    $1.202 \pm 0.003$ &
    $1.209 \pm 0.003$ &
    $1.183 \pm 0.004$ &
    $1.164 \pm 0.005$ \\
  ~~~RMS at phase center & mJy &
    $0.0077$ & $0.0091$ & $0.015$ & $0.017$ & $0.018$ & $0.018$ & $0.015$ \\
  ~~~WISE~0716$-$19 flux density & mJy &
    $0.156 \pm 0.011$ &
    $0.153 \pm 0.013$ &
    $0.105 \pm 0.021$ &
    $0.225 \pm 0.024$ &
    $0.147 \pm 0.026$ &
    $0.279 \pm 0.025$ &
    $0.218 \pm 0.024$ \\[1em]
  7.5 GHz: \\
  ~~~Synthesized beam size & arcsec &
    \nodata &
    \nodata &
    $7.0 \times 2.2$ &
    $6.9 \times 2.3$ &
    $7.3 \times 2.2$ &
    $7.7 \times 2.1$ &
    $7.0 \times 2.5$ \\
  ~~~Calibrator flux density\tablenotemark{a} & mJy &
    \nodata &
    \nodata &
    $0.894 \pm 0.007$ &
    $0.915 \pm 0.004$ &
    $0.933 \pm 0.004$ &
    $0.898 \pm 0.006$ &
    $0.877 \pm 0.007$ \\
  ~~~RMS at phase center & mJy &
    \nodata & \nodata & $0.019$ & $0.015$ & $0.016$ & $0.017$ & $0.014$ \\
  ~~~WISE~0716$-$19 flux density & mJy &
    \nodata &
    \nodata &
    $0.103 \pm 0.027$ &
    $0.132 \pm 0.021$ &
    $0.109 \pm 0.022$ &
    $0.199 \pm 0.025$ &
    $0.067 \pm 0.019$ \\[1em]
  WISE~0716$-$19 spectral index & &
    \nodata &
    \nodata &
    $-0.1 \pm 1.1$ &
    $-1.8 \pm 0.7$ &
    $-1.0 \pm 0.9$ &
    $-1.1 \pm 0.5$ &
    $-3.9 \pm 1.1$ \\
  \enddata
  \tablenotetext{a}{Does not include systematic errors on the absolute flux density scale.}
\end{deluxetable*}

\autoref{t.obs} summarizes our observations and the results of our analysis.
In all cases, the bandpass and flux density calibrator was 3C~147, and the
gain and phase calibrator was the nearby (\apx5\deg\ distant) source
PKS~0733$-$17. A standard continuum wideband correlator setup was used, with
512 channels of 2~MHz width correlated around center frequencies of 5.5~GHz
and 7.5~GHz, the same as used by \citet{kjb+16}. The first two epochs did not
obtain data at the higher frequency. The correlator dump time was 5~s.
Radio-frequency interference was flagged automatically using the
\textsf{aoflagger} tool, which provides post-correlation \citep{odbb+10} and
morphological \citep{ovdgr12} algorithms for identifying interference. After
applying standard calibration techniques in CASA \citep{the.casa}, we imaged
different portions of the data using the CASA imager with 1$''$ square pixels,
128 $w$-projection planes \citep{cgb05}, multi-frequency synthesis
\citep{the.mfs}, and CASA's multi-frequency clean algorithm.

In the images we detect an unresolved source coincident with \obj{0716-19}. In
a stack of all of the data, the position is RA~= 07:16:34.64, Dec.~=
$-$19:00:40.7, with an uncertainty of 0.4~arcsec; this may be compared with
the AllWISE position, RA~= 07:16:34.598, Dec.~= $-$19:00:39.26; and the ATCA
position reported by \citet{kjb+16}, RA~= 07:17:34.6, Dec.~= $-$19:00:40,
where the uncertainties on these are \apx0.1 and \apx1~arcsec, respectively.
The radio positions are consistent with emission from the centroid of the
galaxy.

We measured the source's flux density by least-squares parameter fitting of
the image data and report the results in \autoref{t.obs}, where the flux
density uncertainties are derived from the least-squares covariance matrix.
The minimum and maximum 5.5~GHz flux densities we observe are $0.105 \pm
0.021$ and $0.279 \pm 0.025$~mJy, consistent with the radio transient proposed
by \citet{kjb+16}. We investigated the short-timescale variability of the
radio source using the visibility-based technique described in \citet{wbz13},
finding no evidence of variability on the half-hour time scales of the
individual epochs.

There are two other sources in the VLA field of view that are detectable in
our brief observations. These are found at RA~= 07:16:39.4, Dec.~= $-$18:56:30
and RA~= 07:16:04.0, Dec.~= $-$19:00:16, separated from the pointing center by
1.1 and 1.8 times the half-width at half-power of the VLA primary beam at
5.5~GHz, respectively. Our flux density measurements of these sources vary at
the 10\% and 20\% levels, respectively, which is about twice the level
expected from noise. The variations among the three sources are inconsistent
with an error in the data's overall gain calibration, and extensive checking
of the data reveals no worrisome artifacts. We speculate that the variation we
observe is due to a combination of pointing errors and possibly intrinsic
variability; one of these sources may be the additional radio variable
reported by \citet{kjb+16}.

The typical synthesized beam in the ATCA observations was $10'' \times 2''$
with North-South elongation (S.~Johnston, 2016, priv. comm.), comparable to
that in our VLA observations. Combined with the fact that the source appears
unresolved in both data sets, we infer that a potential systematic flux
density difference due to the ``resolving out'' of flux by interferometers
with different configurations is small. Regardless, any such systematic
difference cannot be responsible for the variation seen in the VLA data set.

\section{Conclusions}
\label{s.conc}

We have pursued three lines of argument against the association between
\thefrb\ and \obj{0716-19} proposed by \citet{kjb+16}. First, the possibility
that \obj{0716-19} is a radio variable was not sufficiently excluded. Second,
the agreement between the DM of the FRB and the redshift of the candidate host
galaxy is not surprising if the host is a randomly-selected radio variable.
Third, the radio light curve of the proposed transient is better explained as
a steady source affected by strong interstellar scintillation, possibly also
showing an AGN flare, than as any of the classes of confirmed extragalactic
radio transients.

We argue that the radio luminosity of \obj{0716-19} indicates that it is
indeed a variable radio source, namely an AGN. Our new data confirm its
variability and show that the galaxy's brightness can reattain the level
attributed to a radio transient by \citet{kjb+16}. The available evidence
therefore cannot support the identification of \obj{0716-19} as the host
galaxy of \thefrb, negating the claimed localization and definitive
cosmological origin of the event.

\acknowledgments

We thank the referees for helpful comments that improved the paper. We thank
Michael Hippke for pointing out the discrepancy between the tabulated and
plotted data in \citet{kjb+16}, Simon Johnston for providing information about
the ATCA data, and Ryan Chornock, Jim Moran, Mark Reid, and Rick Perley for
helpful discussions. The VLA is operated by the National Radio Astronomy
Observatory, a facility of the National Science Foundation operated under
cooperative agreement by Associated Universities, Inc. This work made use of
NASA's Astrophysics Data System; the SIMBAD database, operated at CDS,
Strasbourg, France; and the NASA/IPAC Infrared Science Archive, which is
operated by the Jet Propulsion Laboratory, California Institute of Technology,
under contract with the National Aeronautics and Space Administration.

\facilities{Karl G. Jansky Very Large Array}

\software{CASA, emcee}

\bibliography{\jobname}{}

\end{document}